\documentclass[12pt]{article}
\usepackage{graphics}
\DeclareGraphicsExtensions{.ps,.jpg,.pdf}
\begin{document}
\pagestyle{plain}
\pagenumbering{arabic}
\flushbottom
\begin{titlepage}
\title{Quantum Entanglement dependence on bifurcations and scars in non autonomous systems. The case of Quantum Kicked Top }
\author{George Stamatiou  
\and Demetris P.K. Ghikas
\thanks{Corresponding author: ghikas@physics.upatras.gr, Tel: +302610997460}
\and Department of Physics, University of Patras,\\ Patras 26500, Greece}
\maketitle
\begin{abstract}
Properties related to entanglement in quantum systems, are known to be associated with distinct properties of the corresponding classical systems, as for example stability, integrability and chaos. This means that the detailed topology, both local and global, of the classical phase space may reveal, or influence, the entangling power of the quantum system. As it has been shown in the literature, the bifurcation points, in autonomous dynamical systems, play a crucial role for the onset of entanglement. Similarly, the existence of scars among the  quantum states seems to be a factor in the dynamics of entanglement. Here we study these issues for a non-autonomous system, the Quantum Kicked Top, as a collective model of a multi-qubit system. Using the bifurcation diagram of the corresponding classical limit (the Classical Kicked Top), we analyzed the pair-wise and the bi-partite entanglement of the qubits and their relation to scars, as a function of the critical parameter of the system. We found that the pair-wise entanglement and pair-wise negativity show a strong maximum precisely at the bifurcation points, while the bi-partite entanglement changes slope at these points. We have also investigated the connection between entanglement and the fixed points on the branch of the bifurcation diagram between the two first bifurcation points and we found that the entanglement measures take their extreme values precisely on these points. We conjecture that our results on this behavior of entanglement is generic for many quantum systems with a nonlinear classical analogue.   
\end{abstract}
\end{titlepage}
\newpage
\noindent
\,1.\, INTRODUCTION
\vspace{0.25 cm}\\
\indent Quantum systems with classical analogues can show properties which strongly depend on the detailed topological structure of the classical phase space. One of the oldest and intensively studied properties is the quantum signature of classical chaos, the so called quantum chaos \cite{gutz}. In a broad sense , the last concept is mostly defined in relation to the classical system. Namely, knowing the parameter ranges for which the corresponding classical system is chaotic, one associates quantum structures with characteristic classical behavior in these parameter ranges, as for example the well established eigenvalue and eigenvector distributions \cite{haak,reich}. On the other hand it has been a lot of research activity for a purely quantum mechanical definition of chaos in terms of intrinsic quantum properties, the most characteristic of these being the quantum entanglement \cite{niel}. Indeed it has been shown that the entangling power of the evolution operator and the stability of entanglement are strongly influenced by the properties of the classical regime \cite{bell,mona,fuji,ghos,ghos1,bandy1,bandy2,wang,jac,znid,pine,wein}. Though there are some conflicting results and mostly dependent on specific models, the consensus is that quantum entanglement is an effective signature of quantum chaos. One then may pose the question whether the more detailed structure of the classical phase space may reveal a corresponding structure or behavior of the quantum system. Thus it is natural to be asked if the bifurcation points can signal anything for the quantum regime. After all, the most common scenarios of classical chaos are associated with specific series of bifurcations. The question that has been analyzed in the literature is the behavior of entanglement at the bifurcation points, and it has been shown ,for autonomous systems, that indeed entanglement is sensitive to this classical information \cite{ema,hin1,hin2,nem}. In a different direction, there is an established relationship between classical periodic orbits and localization of wavefunctions, the so called scars \cite{hell,kus,pola,wisn1,wisn2,lee}. The existence of scars is closely related to the chaotic regime, and an immediate question is whether entanglement and scars are related . There exist studies of autonomous systems which answer this question positively \cite{laksh}. In this paper we study these issues for a non autonomous system, the Quantum Kicked Top \cite{haak1,kus1,band}. We use this to model the collective behavior of a system of qubits perturbed externally by discrete kicks. This system has been studied extensively in the literature. It is well understood how the classical and quantum chaos are interrelated, what is the role of periodic orbits, how entanglement production may be quantified and analyzed with this model and finally how entanglement may be used as a signature of quantum chaos. Here first we study the role of bifurcations in the entangling power of the kicked top. We have found that the pair-wise entanglement shows a strong maximum precisely at the bifurcation points. Then we analyzed the relation of the existence of scars  to entanglement properties and we have observed that the bi-partite entanglement as a function of the critical parameter shows a strong change of slope. Finally to have a more complete picture, we investigated the dependence of entanglement on the position of the phase space point with respect to the branch of stable fixed points. We found that , precisely on this branch , between the first two bifurcation points, entanglement becomes extremum. Of course a complete understanding of this behavior would come from an extensive analysis of the relation between entanglement and the stability properties of the points on higher branches of the bifurcation diagram, high values of k, well inside the chaotic regime. In section 2 we introduce the model and review some known facts from the literature. We also present the concept of quantum scars as it is applied in this model. In section 3 we present our calculations and in the final section we discuss our results and present a conjecture about their generality.
\newline
\vspace{0.25 cm}\\ 
 \,2.\,BACKGROUND
\vspace{0.25 cm}\\   
   \,2.1\,Quantum and Classical Kicked Top
\vspace{0.25 cm}\\   
\indent For a system of N qubits we define the collective operators 
\begin{equation}
\hat{J}_{a}\,=\,\sum_{i=1}^{N}\frac{\hat{\sigma}_{ia}}{2}
\end{equation}
with $\{\hat{\sigma}_{ia}\}$ the Pauli matrices. We use as a basis the symmetric Dicke states \cite{dick} $\{|j,m>,m=-j,\cdots,j,j=N/2\}$. This means that the dimension of the space is N+1 and not $2^{N}$, a fact that facilitates the numerical computations. We model the dynamics of this system with the 
 Quantum Kicked Top (QKT) Hamiltonian \cite{kus}. This is the non autonomous quantum system 
\begin{equation}
\hat{H}(t)\,=\,(\hbar p/\tau)\hat{J}_{y}\,+\,(\hbar k/2j)\hat{J}_{z}^{2}\sum_{n=-\infty}^{+\infty}\delta(t-n\tau)
\end{equation}  
The linear term describes a precession around the y axis with angular frequency $p/\tau$. The quadratic perturbation , with perturbation parameter k accounts for periodic kicks in time steps with period $\tau$. The unitary time evolution is given by the Floquet operator
\begin{equation}
\hat{F}\,=\,e^{-i(k/2j)\,\hat{J}_{z}^{2}}e^{-ip\hat{J}_{y}}
\end{equation}  
with the normalization $\hbar=1,\tau=1$. The evolution is a precession around the y axis by an angle p with a  subsequent kick. If the initial state of the system is $|y_{0}>$, after n units of time the state is given by $|y_{n}>\,=\,\hat{F}^{n}|y_{0}>$. The equation
\begin{equation}
\hat{F}|\Phi_{m}>\,=\,e^{ix_{m}}|\Phi_{m}>
\end{equation} 
determines the pseudo-eigenfunctions and the pseudo-eigenvalues of the Floquet operator. These functions provide a complete set for the expansion of the states of the system, namely
\begin{equation}
|y_{n}>\,=\,\sum_{m}e^{inx_{m}}|\Phi_{m}><\Phi_{m}|y_{0}>
\end{equation}
We note here that this dynamics leaves the set of symmetric states invariant. The transition to the corresponding Classical Kicked Top is achieved by first expressing the dynamics of the system in the Heisenberg Picture. The evolution of the angular momentum operators 
\begin{equation}
\hat{J}_{in}\,=\,\hat{F}^{\dag n}\hat{J}_{0}\hat{F}^{n}
\end{equation}
gives the equations
\begin{eqnarray}
\hat{J}_{x}^{'}\,&=&\,\frac{1}{2}(\hat{J}_{x}cosp\,+\,\hat{J}_{z}sinp\,+\,i\hat{J}_{z})e^{i\frac{k}{j}(\hat{J}_{z}cosp-\hat{J}_{x}sinp+\frac{1}{2})}\,+\,h.c \\
\hat{J}_{y}^{'}\,&=&\,\frac{1}{2i}(\hat{J}_{x}cosp\,+\,\hat{J}_{z}sinp\,+\,i\hat{J}_{z})e^{i\frac{k}{j}(\hat{J}_{z}cosp-\hat{J}_{x}sinp+\frac{1}{2})}\,+\,h.c \\
\hat{J}_{z}^{'}\,&=&\,\hat{J}_{z}cosp-\hat{J}_{x}sinp
\end{eqnarray}
In order to define the corresponding Classical Kicked Top one has to take the limit of very large angular momentum j. Defining the new operators $\hat{T}_{a}=\hat{J}_{a}/j \,,\,a\in \{x,y,z\}$ which satisfy $[\hat{T}_{a},\hat{T}_{b}]=\frac{i\epsilon_{abc}}{j}\hat{T}_{c}$, we have that for $j\to \infty$ these operators commute. Then writing $X=T_{x},Y=T_{y},Z=T_{z}$, in this limit , we have the classical equations of motion
\begin{eqnarray}
X'\,&=&\,Re(Xcosp+Xsinp+iY)e^{ik(Zcosp-Xsinp)}\\
Y'\,&=&\,Im(Xcosp+Xsinp+iY)e^{ik(Zcosp-Xsinp)}\\
Z'\,&=&\,-Xsinp+Zcosp
\end{eqnarray} 
For the specific value $p=\pi/2$ the dynamics is given by the equations
\begin{eqnarray}
X'\,&=&\,Zcos(kX)+Ysin(kX)\\
Y'\,&=&\,-Zsin(kX)+Ycos(kX)\\
Z'\,&=&\,-X
\end{eqnarray}
and since we have 
\begin{equation}
X'^{2}+Y'^{2}+Z'^{2}\,=\,X^{2}+Y^{2}+Z^{2}\,=\,1
\end{equation}
the transformation $X=sin\theta cos\phi\,,\,Y=sin \theta sin\phi\,,\,Z=cos\theta$, gives a two dimensional map as the equations of motion for the Classical Kicked Top. This classical dynamics is well analyzed in the literature \cite{haak1,kus1,band}. For small values of k we have islands of regularity which are destroyed as k increases. For intermediate values of k we have a mixed phase space, while for big values we have full chaos. There are two trivial equilibrium points, at the north and south poles of the sphere, defining these points here w.r.t. Y axis. We follow the bifurcation diagram of one of these equilibrium points analyzing the behavior of the corresponding quantum system at the bifurcation point. In a more complete analysis , we plan to track the signatures of many bifurcations points and picture the overall significance of the bifurcation landscape for the quantum behavior. There are generic scenarios of the onset of chaos, and these are well studied and understood, as well as cases of  systems with singular behavior. We consider the study of the quantum signature of these classical portraits as very important. In this paper we are only taking the very first step of such an analysis.         
\vspace{0.25 cm}\\
\,2.2\,Quantum Scars.
\vspace{0.25 cm}\\
\indent In the semi-classical approximation theory of Quantum Mechanics the classical periodic orbits play a fundamental role \cite{gutz}. It has been proved that the fluctuating part of the density of states can be written as a sum of terms over the classical periodic orbits, and that the existence of these orbits influences the energy spectrum of the corresponding quantum system. But one of the most impressive and unexpected results is the existence of scars \cite{hell}. These are states which are strongly localized in the phase space at points which are directly related to the classical periodic orbits. These Quantum Scars (QS) do not exist for every periodic orbit and not every quantum state is related to any periodic orbit. But it is interesting to investigate how the existence of QS affects the quantum properties of the system. For non-autonomous systems, like the QKT, scars can be found among the pseudo-eigenstates of the Floquet operator. It has been proved that scars play a significant role in the time evolution of a coherent state which starts from a point of the periodic orbit \cite{hell,kus,pola,wisn1,wisn2,lee}. In this paper we analyze the relation of scars to the entanglement. We base our analysis on some known results. 
\newline
\indent Let $|\gamma>=|\theta,\phi>$ be a given state of the QKT where $|\theta,\phi>$ is the spin coherent state defined by \cite{arre}, $\{|\theta,\phi>=R(\theta,\phi)|j,j> ; -\pi\leq\phi \leq\pi, 0\leq \theta \leq\pi\}$, with 
\begin{equation}
R(\theta,\phi)\,=\,exp\{i\theta[J_{x}sin\phi - J_{y}cos\phi]\}
\end{equation}
Then for the nth iterate of the Floquet operator we have $|\gamma'> = \hat{F}^{n}|\gamma>$. Here $\gamma'$ is the coherent state parameter given by the classical dynamics $\gamma ' = \underbrace {T(\cdots T(T(\gamma)))}_{n}$, where T is the appropriate classical limit of the Floquet map. Thus for a classical periodic orbit of order K we have   $\gamma_{0}  = \underbrace {T(\cdots T(T(\gamma_{0})))}_{K}$,  \cite{haak1}. Then for the M-step propagator $f_{M}=<\gamma_{0}|\hat{F}^{M}|\gamma_{0}>$ we have, after expanding in the pseudo-eigenfucntion basis 
\begin{equation}
f_{M}\,=\,\sum_{n=1}^{N}H_{n}(\gamma_{0})e^{iM x_{n}}
\end{equation}
where N is the dimensionality of the system and $H_{n}=|<\Phi_{n}|\gamma_{0}>|^{2}$ the Husimi representation \cite{kus} of the nth pseudo-eigenfunction $\Phi_{n}$. The importance of this quantity is that its Fourier Transform
\begin{equation}
\hat{f}_{\omega}\,=\,\sum_{M=-L}^{L}e^{-i\omega M}f_{M}\,=\,\sum_{n=1}^{N}H_{n}(\gamma_{0})\sum_{M=-L}^{L}e^{iM(\phi_{n}-\omega)}
\end{equation}
contains the information for the quantum scars. That is, when $\omega$ is close to a pseudo-eigenenergy, the spectral density is dominated by the overlap function of the corresponding pseudo-eigenfunction if the latter is associated to a scar. Thus we locate the QS that corresponds to the periodic orbit 
$\Gamma =(\gamma_{0},\gamma_{1},\cdots ,\gamma_{M-1})$ as a maximum in $\hat{f}_{\omega}$. Stated differently, the eigenfunction associated with the energy at the peak is scarred by this periodic orbit \cite{kus}. We refer to the literature for more details on scars and their properties.  
\vspace{0.5 cm}\\
\,2.3\,Entangling Measures.
\vspace{0.25 cm}\\
\indent Our main objective is to study the pair-wise and the bi-partite entaglement in a system of N qubits which is subjected to a time-dependent external perturbation and analyze the interrelation of these quantities with the existence of bifurcation points of the corresponding classical system and the role of quantum scars. As we said above, we model this collective behavior as Classical and Quantum Kicked Tops, the properties of which are very extensively analyzed in the literature. We quantify the pair-wise entanglement with both the entanglement of formation and negativity and the bi-partite entanglement with the linear entropy. More specifically, let $\hat{\rho}$ be the density matrix of the system and ${|i_{1}i_{2}\cdots i_{N}>}\,i_{k}\in \{0,1\}$ the computational basis. Let $\hat{\rho}_{k}=tr_{ALL\ne k}(\hat{\rho})$ and  $\hat{\rho}_{km}=tr_{ALL\ne k,m}(\hat{\rho})$ be the one qubit and two qubits reduced density matrices. We use three measures of entanglement. The bi-partite entanglement $Q \equiv 2-\frac{2}{N}\sum_{i}tr(\hat{\rho}_{i}^{2})$, the Negativity $N_{p}= N(\hat{\rho}_{km})$ and the Entropy of Formation $E_{p}= E_{formation}(\hat{\rho}_{km})$. Negativity is defined as follows : For a generally mixed state $\hat{\rho}$, $N(\hat{\rho)}= \sum_{j}max\{0,-\mu_{j}\}$ where $\{\mu_{j}\}$ are the eigenvalues of the partial transpose of $\hat{\rho}$. Entanglement of Formation is defined as $E_{f}=H(\frac{1+\sqrt{1-C^{2}}}{2})$ where H(x) is the Shannon entropy function and C the concurrence defined by $C=max\{0,\lambda_{1}-\lambda_{2}-\lambda_{3}-\lambda_{4}\}$ and the quantities  $\lambda_{j}$ are , in decreasing order , the square roots of the eigenvalues of the matrix $\hat{\rho}_{12}(\sigma_{1y}\otimes \sigma_{2y})\hat{\rho}_{12}^{*}(\sigma_{1y}\otimes \sigma_{2y}))$. We refer to \cite{ben,woo,per,horo,zycz,vid} for more details on the properties of these quantities. In the Discussion we comment on the use of Generalized Entanglement \cite{barn1,barn2,wein}.  
\vspace{0.5 cm}\\ 
\,3.\,THE RESULTS
\vspace{0.25 cm}\\  
\,3.1\,The Classical Picture.
\vspace{0.25 cm}\\     
We first present a series of portraits of the Classical Kicked Top for different  values of the parameter k and the bifurcation  diagrams for the angles $\theta_{0}$ and $\phi_{0}$ of the equilibrium point of the south pole.
\noindent
\newline
\indent In Figure 1 we reproduce the phase space portrait for four different values of the coupling constant parameter k, with one value at the bifurcation point k=2. We present these four figures to point to the way the classical phase space picture is evolving with  the increase of parameter k. Since we are not studying here the classical behavior of this system, we are not concerned with the small structures evident in the figures as we approach the chaotic regime. It is , of course, very interesting to study the relation of these details to the quantal properties of the associated system. We only note here, that in the limit of fully developed chaos, there exist in general unstable periodic orbits which we expect to play a role in the behavior of the quantum system. In Figure 2 we plot the bifurcation diagrams for the angles $\theta$ and $\phi$ which start from the equilibrium point in the south pole. We note that the fuzziness of the bifurcation diagrams after the point k=2 and before the area of complete chaos is due to numerical inaccuracies. In this figure, the first bifurcation point is more than evident. In our results , below, for the entanglement, the signature of the existence of this point in the behavior of the quantum system is clearly seen. But it is not yet clear to us how to pin point the onset of the band. Perhaps, as it is commented in the Discussion, other measures, like Generelaized Entanglement may offer a better picture. 
\noindent
\vspace{0.25 cm}\\
\,3.2\,Bifurcation Points and Entanglement of Scars.
\vspace{0.25 cm}\\   
\indent As it was described in the Introduction, scars are quantum states which are associated to classical periodic orbits and appear as prominent structures in phase space through their Husimi representation. In order to relate the entanglement of the scars to the bifurcation point at k=2, we first identify the scar located at the south pole for each value of k, from k=0 to k=6.8. Then for each value of k we compute for the corresponding scar the pairwise negativity, the pairwise entanglement of formation and the bipartite entanglement. We did the calculations for j=35, that is N=70 qubits. In Figure 3 we plot the scars and the associated spectrum corresponding to the values of k in Figure 1.
In this figure it is seeing that as k increases more peaks are present, and as k approaches the region of full chaos the peaks multiply greatly. We identify the highest peak with the scar that we study. The corresponding plot of the Husimi function shows a progressive delocalization. In Figure 4 we plot the three different entanglement measures, pair-wise negativity, pairwise entanglement and bipartite entanglement for N=50 and N=110 qubits. First we observe that in all plots the signature of the bifurcation point at k=2 is more than evident. Second, there is not any dramatic change with the increase of the number of qubits. In this figure we see that though the entanglement is associated clearly with the scarred function at the bifurcation point, this association is not strict, and it shows a relatively wide span of values where the entanglement is appreciable. It seems that this is related to the progressive delocalization of the scarred state.  
\vspace{0.25 cm}\\ 
\,3.3\,Entanglement and Periodic Orbits.
\vspace{0.25 cm}\\     
In Figure 5 we plot the pair-wise entanglement of formation and the bi-partite entanglement for time evolved states. Specifically, we start with two coherent states located at the south and north poles respectively, and we compute their entanglement after 500 kicks. We repeat this for a large number of k values from k=0 to k=6.8. We observe again the trace of the classical bifurcation point at k=2. There is not any difference between south and north poles and not a strong dependence on the number of qubits for N in the order of 10. It is interesting to note that classically the two poles behave differently at the bifurcation points \cite{haak1}. Here , again, though the value of k=2 is evidently seeing to be prominent, there is not a sharp peak. It seems that the bifurcation point is smoothly felt by the periodic orbit starting from the equilibrium points. A more extensive study is probably needed to understand   the influence of the local structure of the classical phase space on the quantum behavior.   
\vspace{0.25 cm}\\ 
\,3.4\,Entanglement along the branch of stable points.
\vspace{0.25 cm}\\     
To have a more complete picture of the role of the classical stable fixed points we analyzed the dependence of entanglement on the location of the fixed points w.r.t. the branch of the stable points between the values k=2 and k=12.73, the latter corresponding to the second bifurcation point. We computed the bipartite entanglement and the pairwise negativity for various values of k in the above interval. For the k=12.73 the fixed point is $(\theta,\phi)=(2.32,0.38)$, and we computed the dependence of entanglement on $\theta$ and $\phi$. We found that, at precisely this fixed point, the entanglemnt is minimum. In Figure 6 third row, we plot the bipartite entanglement vs either $\theta$ or $\phi$. In the same figure, first and second rows we plot the results for k=2.2 and k=2.4. For these cases the corresponding values for the fixed points are  $(\theta,\phi)=(1.9,1.2),(2.1,1.0)$. We keep $\theta$ fixed and we plot both the bipartite entanglement and the pairwise negativity vs $\phi$. Again we observe that the fixed points give minimum entanglement. We note that the same behaviour is observed for all values of k in the interval [2,12.73].  
\vspace{0.25 cm}\\ 
\,4.\,DISCUSSION
\vspace{0.25 cm}\\ 
\indent In this paper we posed the question of the relation of the local structures of classical phase space to the purely quantum mechanical property of entanglement. Generally, it is well established that, both qualitatively and quantitatively, entanglement and its fragility, namely the properties of the decoherence process, strongly depend on the classical regime. That is, the presence of classical structures, in the corresponding classical system, can be traced in the properties of the quantum system. But the onset of classical chaos is generically related to bifurcations from equilibrium points, and the scenarios of chaos are pictured in connection to local or global changes of the phase space portrait of the system. It is generally expected that a global change of the classical landscape can be felt by the quantum system. But it seems very natural to ask whether a local property, that is a single bifurcation point can be traced in the quantum properties. It is this question that we have tried to answer here using scars and periodic orbits as "tools" of the analysis. We have used the Classical and Quantum Kicked Tops because they have been well studied and understood. We use the top as a collective model of N qubits. We applied three measures of entanglement , pair-wise negativity, pair-wise entanglement of formation and bi-partite entanglement. We traced the behavior of the system for many values of the coupling parameter k, from k=0 to k=6.8. As it is known for k=2 we have the first bifurcation point of the equilibrioum at the south and north poles. First we presented four portraits of the phase space for characteristic k values and the bifurcation diagrams of the south pole. Then for each value of k in the above range we locate the scarring state and compute the entanglement. The scars along with the associated spectra are plotted for the selected four values of k. Finally we propagated two coherent states , initially located at the south and north poles and we computed their entanglement after 500 kicks. In all plots of the measures of entaglement the bifurcation point at k=2 is more than evident. We conjecture that this is not a non-generic behavior. That is we expect that the bifurcation points are intimately related to the quantum structures. But , as it happens in the case of scars, we cannot tell whether all bifurcation points play this role and moreover whether the sensitivity of entanglement can always be related to a bifurcation phenomenon in the classical regime. To establish a more complete picture, at this stage of our investigation, we made an analysis of the dependence of entanglement on the propery of stability of the fixed points. On the branch of stable fixed points in the bifurcation diagram between k=2 and k=12.73 (first and second bifurcation points), for various values of k, we computed the dependence of entanglement on the position of the point in phase space. We observed that when the point crosses the branch the entanglement becomes minimum. We ploted the cases for k=2.2, 2.4 and 12.73. The behavior is similar for other values of k in this interval. Given the fact that the bifurcation diagram is very complicated it would be interesting to see whether for very high values of k, we have a new pattern on the entanglement dependence. We are currently investigating these issues. We end with few comments on the use of Generalized Entanglement (GE) \cite{barn1,barn2,wein}. The static or dynamic signatures of quantum chaos, like eigenvalue distributions or fidelity, and entanglement generation, with the associate entanglement power of the evolution operator are widely used, but depend strongly on the basis employed. But more importantly, they depend on the partition of the given system into subsystems, or the embedding into a larger system. It is this practice which we follow here using the pair-wise and the bi-partite entanglent and the negativity. On the other hand the GE depends on a selected subalgebra of observables. The regime is characterized by sums of averages of these operators. These averages are sensitive detectors of the global classical phase space, and for this reason GE is considered presently as the most effective signature of quantum chaos. In the present paper we were interested on two critical details of the classical dynamics, namely the role of bifurcation points and of general periodic solutions in the entanglement affected by the dynamics on the subystems. Thus from the start our interest is on subsystems and their combined state, and according to \cite{barn2} our bi-partite entanglement measure is actually related to GE.  Of course we consider the question of the dependence of GE on the local properties of the phase space as very important, and we are looking into this issue.  

\begin{thebibliography}{99}
\bibitem{gutz}
M.C. Gutzwill, \textit{Chaos in Classical and Quantum Mechanics}, Springer-Verlag, New York, 1990.
\bibitem{haak}         
F. Haake, \textit{Quantum Signatures of Chaos},Springer-Verlag, Berlin, 1991.
\bibitem{reich}
L.E. Reichl,\textit{The Transition to Chaos,Conservative Classical Systems and Quantum Manifestations}, Springer-Verlag,New York, 2004. 
\bibitem{niel}
M.A. Nielsen and I.L. Chuang, {\itshape Quantum Computation and Quantum Information\/}, Cambridge University Press, Cambridge, UK, 2000
\bibitem{furu}
K. Furuya, M.C. Nemes, G.Q. Pellegrino, Phys. Rev. Lett. \begin{bfseries}25\end{bfseries}, 5524 (1998).
\bibitem{bell}
N.F. Bell, R.F. Sawyer, R. R. Volkas, Y.Y.Y. Wong Phys. Rev A \begin{bfseries}65\end{bfseries}, 042328 (2002) 
\bibitem{bandy1}
J.N. Bandyopadhyay, A. Lakshminarayan, Phys. Rev. Lett. \begin{bfseries}89\end{bfseries} 060402 (2002).
\bibitem{bandy2}
J.N. Bandyopadhyay, A. Lakshminarayan, Phys. Rev. E \begin{bfseries}69\end{bfseries} 016201 (2004).
\bibitem{mona}
C. Mejia-Monasterio, G. Benenti, G. G. Carlo, G. Casati Phys. Rev A \begin{bfseries}71\end{bfseries}, 062324 (2005)
\bibitem{fuji}
H. Fujisaki, Phys. Rev A \begin{bfseries}70\end{bfseries}, 012313 (2004)
\bibitem{ghos}
S. Ghose, B.C. Sanders, Phys. Rev A \begin{bfseries}70\end{bfseries}, 062315 (2004)
\bibitem{ghos1}
S. Ghosh, P.M. Alsing, I.H. Deutsch, Phys. Rev E \begin{bfseries}64\end{bfseries}, 056119 (2001).
\bibitem{wang}
X. Wang, S. Ghose, B.C. Sanders, B. Hu, e-print quant-ph/0312047.
\bibitem{jac}
Ph. Jacquod, Phys. Rev. Lett. \begin{bfseries}92\end{bfseries} 150403 (2004).
\bibitem{znid}
M. Znidaric, T. Prosen, Phys. Rev A \begin{bfseries}71\end{bfseries}, 032103 (2005).
\bibitem{pine}
C. Pineda, T.H. Seligman, e-print quant-ph/0503177.
\bibitem{wein}
Yaakov S. Weinstein, Lorentza Viola, quant-ph/0603071(2006).
\bibitem{ema}
Clive Emary, Neil Lambert,Tobias Brandes, PHYS. REV. A \begin{bfseries}71\end{bfseries}, 062302, 2005.
\bibitem{hin1}
A.P. Hines, G.J. Milburn, R.H. McKentzie, quant-ph/0308165(2003). 
\bibitem{hin2}
A.P. Hines, C.M. Dawson, R.H. McKenzie, G.J. Milburn, PHYS. REV. A \begin{bfseries}70\end{bfseries}, 022303 (2004).
\bibitem{nem}
M.C. Nemes, K. Furuya, G.Q. Pellegrino, A.C. Oliveira, Mauricio Reis, L. Sanz, quant-ph/0507026 (2005).
\bibitem{hell}
Eric J. Heller, Phys. Rev. Lett. \begin{bfseries}53\end{bfseries}, 1515 (1984).
\bibitem{kus}
Marek Kus, Jakob Zakrzewski, Karol Zyczkowski, Phys. Rev. A \begin{bfseries}43\end{bfseries},4244 (1991).
\bibitem{pola}
G.G. de Polavieja, F. Borondo, R.M. Benito, Phys. Rev. Lett. \begin{bfseries}73\end{bfseries},1613 (1994).
\bibitem{wisn1}
D.A. Wisniacki, F. Borondo, E. Vergini, R.M. Benito, Phys. Rev. E \begin{bfseries}62\end{bfseries}, R7583 (2000).
\bibitem{wisn2}
.A. Wisniacki, F. Borondo, E. Vergini, R.M. Benito,nlin.CD/0103031.
\bibitem{lee}
Soo-Young Lee, Stephen C. Creagh, nlin.CD/0304018.
\bibitem{haak1}
F. Haake, M. Kus, R. Scharf, Z. Phys. B, Condensed Matter \begin{bfseries}65\end{bfseries}, 381-395 (1987).
\bibitem{kus1}
Marek Kus, Fritz Haake, Bruno Eckhardt, Z. Phys. B, Condensed Matter \begin{bfseries}92\end{bfseries}, 221-233,(1993).
\bibitem{band}
J.N. Bandyopadhyay, A. Lakshminarayan, Phys. Rev. E \begin{bfseries}69\end{bfseries}, 016201 (2004).
\bibitem{laksh}
Aral Lakshminarayan, Phys. Rev. E \begin{bfseries}64\end{bfseries},036207 (2001).
\bibitem{arre}
F.T. Arrechi, E. Courtens, R. Gilmore, and H. Thomas, Phys. Rev. A \begin{bfseries}36\end{bfseries},5543 (1987)
\bibitem{dick}
R.H. Diche, Phys. Rev. \begin{bfseries}93\end{bfseries}, 99 (1954)
\bibitem{barn1}
H. Barnum, E. Knill, G. Ortiz, and L. Viola, Phys. Rev. A \begin{bfseries}68\end{bfseries},032308 (2003)
\bibitem{barn2}
H. Barnum, E. Knill, G. Ortiz, R. Somma, and L. Viola, Phys. Rev. Lett. \begin{bfseries}92\end{bfseries},107902 (2004)
\bibitem{ben}
C.H. Benett, D.P. DiVincenzo,J.A. Smolin and  W.K. Wooters  Phys. Rev. A \begin{bfseries}54\end{bfseries},3824 (1996)
\bibitem{woo}
W.K. Wooters, Phys. Rev. Lett. \begin{bfseries}80\end{bfseries},2245 (1998)
\bibitem{per}
A. Peres, Phys. Rev. Lett. \begin{bfseries}77\end{bfseries},1413 (1996)
\bibitem{zycz}
K. Zyczkowski, P. Horodecki, A Sampera, and M. Lewenstein, Phys. Rev. A \begin{bfseries}58\end{bfseries},883 (1998)
\bibitem{vid}
G. Vidal,and R.F. Werenr, Phys. Rev. A \begin{bfseries}65\end{bfseries},032314 (2002)
\bibitem{horo}
M. Horodecki, P. Horodecki, and R. Horodecki, Phys. Lett. A \begin{bfseries}223\end{bfseries},1 (1196)
\end{thebibliography}
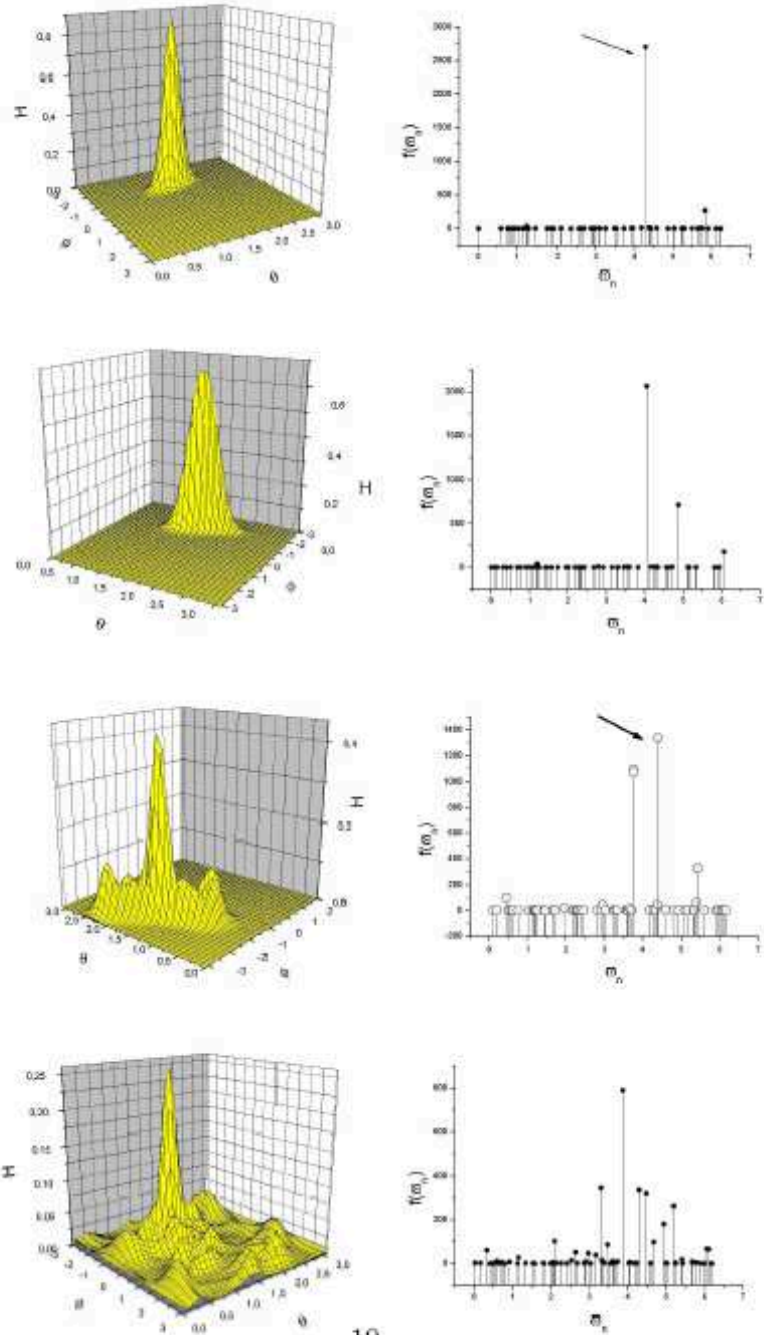
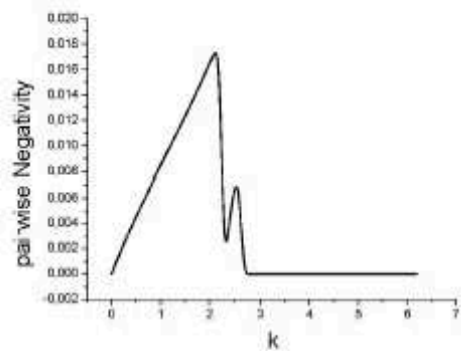
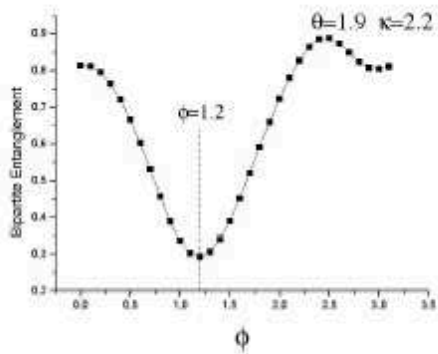
\end{document}